\documentclass{article}
\usepackage{ijcai27}

\usepackage{times}
\usepackage{soul}
\usepackage{url}
\usepackage[hidelinks]{hyperref}
\usepackage[utf8]{inputenc}
\usepackage[small]{caption}
\usepackage{graphicx}
\usepackage{amsmath}
\usepackage{amsthm}
\usepackage{booktabs}
\usepackage{algorithm}
\usepackage{algorithmic}
\usepackage[switch]{lineno}

\usepackage{xcolor}
\usepackage{tikz}
\usetikzlibrary{arrows.meta,positioning,calc}
\usepackage{multirow}
\usepackage{array}

\makeatletter
\def\@citex[#1]#2{%
  \def\@citea{}%
  \@cite{\@for\@citeb:=#2\do
    {\@citea\def\@citea{,\penalty\@citeseppen\ }%
     \if@filesw\immediate\write\@auxout{\string\citation{\@citeb}}\fi
     \@ifundefined{b@\@citeb}{{\bf ?}\@warning
       {Citation `\@citeb' on page \thepage \space undefined}}%
     {\csname b@\@citeb\endcsname}}}{#1}}
\makeatother

\definecolor{leadercol}{HTML}{2A78D6}   
\definecolor{readercol}{HTML}{EB6834}   
\definecolor{ink}{HTML}{0B0B0B}
\definecolor{inkmuted}{HTML}{52514E}

\newcommand{\dthree}{d3}
\newcommand{\tool}[1]{\textit{#1}}

\title{Delegate or Pay: LLM Agents Navigating Knowledge Graphs with Decision Models}

\author{
    Riccardo Terrenzi\And
    Serkan Ayvaz
    \affiliations
    Centre for Industrial Software, University of Southern Denmark, S{\o}nderborg, Denmark
    \emails
    \{rite, seay\}@mmmi.sdu.dk
}

\begin{document}

\maketitle

\begin{abstract}
In agentic knowledge graph question answering (KGQA), an LLM agent finds the answer to a question by navigating the graph step by step, reading the relations of each entity it visits and the facts they lead to.
Since the LLM both directs the search and reads everything it returns, it carries the whole load on its own.
Yet much of this load results in wasted tokens, because many entities have hundreds of relations, and choosing the ones that fit the question is often too trivial to deserve an expensive LLM.
Existing approaches still leave these choices to the LLM and only control how much it reads, either pruning the candidates at the risk of discarding the answer, or showing them all and paying for every token.
We propose a third option and delegate the choices to a decision model.
Such a model is well suited to this task, as it understands the context like an LLM and chooses among given options, such as which relation fits or whether a row meets a condition, at a fraction of the cost.
In our agent, an LLM navigator plans the search through four tools and sees only a compact summary of each result, while the decision model behind the tools does the reading.
Evaluations on four Freebase benchmarks indicated that our agent achieves performance comparable to Search-on-Graph, the strongest training-free agent, while its LLM reads 4 to 6 times fewer tokens.
The larger the neighborhoods it explores, the larger the saving.
\end{abstract}

\section{Introduction}
\label{sec:intro}

\begin{figure*}[t]
  \centering
\begin{tikzpicture}[
    font=\scriptsize, x=1cm, y=1cm,
    sees/.style={align=left, inner sep=3pt, anchor=north west, text width=3.04cm,
                 minimum height=1.05cm},
    calls/.style={align=left, inner sep=3pt, anchor=north west, text width=3.04cm,
                  minimum height=0.95cm, fill=leadercol!12},
    wsmall/.style={font=\fontsize{5.5}{6.5}\selectfont, text width=1.32cm, minimum height=0.95cm},
    wnode/.style={draw=inkmuted, fill=white, rounded corners=2pt, thick, align=left,
                  inner sep=2.5pt, anchor=west, minimum height=0.8cm},
    num/.style={circle, fill=leadercol, text=white, font=\tiny\bfseries, inner sep=0pt,
                minimum size=8pt},
    rec/.style={draw=readercol!40, fill=readercol!8, minimum size=0.14cm, inner sep=0pt},
    keep/.style={draw=readercol, fill=readercol!60, thick, minimum size=0.14cm, inner sep=0pt},
    call/.style={-{Stealth[length=3.5pt]}, thick, leadercol},
    back/.style={-{Stealth[length=3.5pt]}, thick, inkmuted, rounded corners=3pt},
    ask/.style={-{Stealth[length=3.5pt]}, thick, readercol},
    lbl/.style={font=\tiny, inner sep=1pt, align=left, anchor=west},
    head/.style={font=\tiny\bfseries, inner sep=0pt, anchor=west, text=readercol!75!black},
    muted/.style={text=inkmuted},
    tag/.style={font=\fontsize{5}{6}\selectfont\bfseries},
    band/.style={rounded corners=3pt},
    side/.style={rotate=90, font=\tiny\bfseries, text=white, inner sep=0pt},
  ]
  \def\ca{2.6} \def\cb{6.3} \def\cc{10.0} \def\cd{13.7} \def\cw{3.25}

  \node[anchor=west, inner sep=0pt] at (0, 7.65)
    {\footnotesize\textbf{Question:} \emph{who is the head coach of inter milan?}\enspace
     {\color{inkmuted}(WebQSP test, WebQTest-306)}};
  \fill[band, leadercol!5]  (0, 5.0)  rectangle (17.75, 7.35);
  \fill[band, ink!4]        (0, 3.15) rectangle (17.75, 4.35);
  \fill[band, readercol!5]  (0, 0)    rectangle (17.75, 2.5);
  \fill[band, leadercol]    (0, 5.0)  rectangle (0.38, 7.35);
  \fill[band, inkmuted]     (0, 3.15) rectangle (0.38, 4.35);
  \fill[band, readercol]    (0, 0)    rectangle (0.38, 2.5);
  \node[side] at (0.19, 6.175) {Navigator (LLM)};
  \node[side] at (0.19, 3.75)  {Workspace};
  \node[side] at (0.19, 1.25)  {Decision model};

  \foreach \c/\k/\s/\act in {
      \ca/1/{the question and its topic\\ entity \tool{a1}: Inter Milan}/
           {\tool{step}(\tool{a1}, ``head coach'')\\ \strut},
      \cb/2/{\tool{c1}: \textbf{73} records, with\\ fields manager, team, from, to}/
           {\tool{filter}(\tool{c1}, ``to is empty\\ \phantom{\tool{filter}(}or latest'')},
      \cc/3/{\tool{c2}: \textbf{3} records, few enough\\ to be shown in full}/
           {\tool{select}(\tool{c2}, ``latest\\ \phantom{\tool{select}(}from date'')},
      \cd/4/{\textbf{1} record: Mazzarri, from\\ 2013-05-24, evidence \tool{e73}}/
           {\tool{finish}(Walter Mazzarri;\\ \phantom{\tool{finish}(}\tool{e73}, \tool{e291}, \tool{e146})}}{
    \node[sees] (p\k) at (\c, 7.21) {\hspace*{8pt}{\fontsize{5}{6}\selectfont\bfseries\color{inkmuted}SEES}\\ \s};
    \node[calls] (q\k) at (p\k.south west) {\hspace*{8pt}{\fontsize{5}{6}\selectfont\bfseries\color{leadercol}CALLS}\\ \act};
    \draw[leadercol, thick, rounded corners=2pt] (p\k.north west) rectangle (q\k.south east);
    \node[num] at ([xshift=1pt, yshift=-1pt]p\k.north west) {\k};
  }

  \node[wnode, text width=1.4cm] (w0) at (0.55, 3.75)
    {\tool{a1}:\\ Inter Milan\\ {\fontsize{5}{6}\selectfont\color{inkmuted}topic entity}};
  \foreach \k/\c/\l/\r in {
      1/\ca/{lists the 199\\ relations of \tool{a1}}/{\tool{c1}: 73 records,\\ via the chosen\\ relation},
      2/\cb/{attaches the\\ fields to the\\ 73 rows of \tool{c1}}/{\tool{c2}: the 3\\ records kept},
      3/\cc/{offers the\\ 3 rows of \tool{c2}}/{the chosen\\ record, \tool{e73}}}{
    \node[wnode, wsmall] (l\k) at (\c, 3.75) {\l};
    \node[wnode, wsmall] (w\k) at (\c+1.72, 3.75) {\r};
  }
  \node[wnode, wsmall] (l4) at (\cd, 3.75) {checks the\\ cited evidence};
  \node[wnode, wsmall, draw=ink, fill=leadercol!8] (w4) at (\cd+1.72, 3.75)
    {answer:\\ \textbf{Walter}\\ \textbf{Mazzarri}};
  \draw[back] (l4.east) -- (w4.west);

  \foreach \k in {1, 2, 3, 4}
    \draw[call] (l\k.north |- q\k.south) -- (l\k.north);
  \foreach \k in {1, 2, 3}{
    \draw[ask] (l\k.south) -- (l\k.south |- 0, 2.52);
    \draw[ask] (w\k.south |- 0, 2.52) -- (w\k.south);
  }
  \foreach \w/\p/\x in {w0/p1/\ca-0.225, w1/p2/\cb-0.225, w2/p3/\cc-0.225, w3/p4/\cd-0.225}
    \draw[back] (\w.east) -| (\x, 6.68) -- (\x+0.225, 6.68);
  \node[lbl, text=leadercol] at ([xshift=2pt]l1.north |- 0, 4.7) {call};
  \node[lbl, text=inkmuted, anchor=east] at (\cb-0.27, 4.7) {result: only its shape};
  \node[lbl, text=readercol!75!black] at ([xshift=2pt]l1.south |- 0, 2.83) {asks};
  \node[lbl, text=readercol!75!black] at ([xshift=2pt]w1.south |- 0, 2.83) {probabilities};

  \node[head] at (\ca, 2.3) {ranks the relations for ``head coach''};
  \foreach \y/\r/\p/\pl/\hl in {
      2.0/{football\_team.manager}/0.48/.48/1,
      1.72/{sports\_team.coaches}/0.39/.39/0,
      1.44/{sports\_team\_extra.managers}/0.06/.06/0,
      1.16/{organization.leadership}/0.02/.02/0,
      0.88/{sports\_team.roster}/0.004/{$<$.01}/0}{
    \ifnum\hl=1
      \fill[readercol!18, rounded corners=1pt] (\ca-0.05, \y-0.13) rectangle (\ca+\cw+0.05, \y+0.13);
    \fi
    \node[anchor=west, inner sep=0pt, font=\fontsize{5}{6}\selectfont\ttfamily] at (\ca, \y) {\r};
    \fill[readercol] (\ca+2.8, \y-0.05) rectangle ++({\p*0.5}, 0.1);
    \node[anchor=west, inner sep=1pt, font=\tiny] at (\ca+2.8+\p*0.5, \y) {\pl};
  }
  \node[lbl, muted] at (\ca-0.03, 0.55) {+ 194 more relations, each below .01};

  \node[head] at (\cb, 2.3) {judges each of the 73 records, keeps 3};
  \def\gs{0.2}
  \foreach \i in {0,...,72}{
    \pgfmathtruncatemacro\cl{mod(\i,15)}
    \pgfmathtruncatemacro\rw{div(\i,15)}
    \pgfmathtruncatemacro\k{(\i==69) || (\i==70) || (\i==72)}
    \ifnum\k=1 \node[keep] at (\cb+0.07+\cl*\gs, 2.0-\rw*\gs) {};
    \else      \node[rec]  at (\cb+0.07+\cl*\gs, 2.0-\rw*\gs) {};
    \fi
  }
  \node[lbl, muted] at (\cb-0.03, 1.0) {by start date, Fossati (1909) to Mazzarri (2013)};
  \foreach \y/\t/\p/\pl in {0.7/{Gasperini, 2011--12}/0.51/.51, 0.45/{Ranieri, 2011--12}/0.63/.63,
                            0.2/{Mazzarri, 2013--}/0.96/.96}{
    \node[anchor=west, inner sep=0pt, font=\tiny] at (\cb, \y) {\t};
    \fill[readercol] (\cb+1.75, \y-0.05) rectangle ++({\p*1.2}, 0.1);
    \node[anchor=west, inner sep=1pt, font=\tiny] at (\cb+1.75+\p*1.2, \y) {\pl};
  }

  \node[head] at (\cc, 2.3) {chooses one of the 3 records};
  \foreach \y/\t/\p/\pl/\hl in {2.0/{Mazzarri, 2013--}/0.93/.93/1,
                                1.72/{Gasperini, 2011--12}/0.004/{$<$.01}/0,
                                1.44/{Ranieri, 2011--12}/0.004/{$<$.01}/0}{
    \ifnum\hl=1
      \fill[readercol!18, rounded corners=1pt] (\cc-0.05, \y-0.13) rectangle (\cc+\cw+0.05, \y+0.13);
    \fi
    \node[anchor=west, inner sep=0pt, font=\tiny] at (\cc, \y) {\t};
    \fill[readercol] (\cc+1.75, \y-0.05) rectangle ++({\p*1.2}, 0.1);
    \node[anchor=west, inner sep=1pt, font=\tiny] at (\cc+1.75+\p*1.2, \y) {\pl};
  }

  \node[lbl, muted] at (\cd-0.03, 2.3) {no decision-model request: the\\ workspace checks the cited evidence};
  \node[head, text=ink] at (\cd, 1.62) {Input tokens of the question (to scale)};
  \node[lbl] at (\cd-0.03, 1.3) {Navigator: 8.1k in 4 prompts (largest 2.7k)};
  \fill[leadercol] (\cd, 1.0) rectangle ++(8.115/7, 0.13);
  \node[lbl] at (\cd-0.03, 0.7) {Decision model: 20.5k in 4 requests};
  \fill[readercol] (\cd, 0.4) rectangle ++(20.494/7, 0.13);
\end{tikzpicture}
  \caption{The agent on a WebQSP test question (WebQTest-306), one column per step. Each card of the navigator (top) shows what the navigator sees, the question and topic entity \tool{a1} or the result of the previous step, and the tool it then calls. The workspace (middle) turns the call into options (left box), asks the decision model (bottom) to choose among them, and builds the next object from its answer (right box); only a compact summary of this object returns to the navigator (grey arrows), as input to its next call. The decision model ranks the 199 relations of Inter Milan, judges each of the 73 manager records against the condition and keeps 3, then selects the latest; \tool{finish} needs no decision-model request, since the workspace checks that the answer occurs in the cited evidence. The navigator never sees the relation list or the 73 records. Bars show the decision model's probabilities. Relation names are shown without their domain prefix (such as \tool{soccer.}).}
  \label{fig:overview}
\end{figure*}
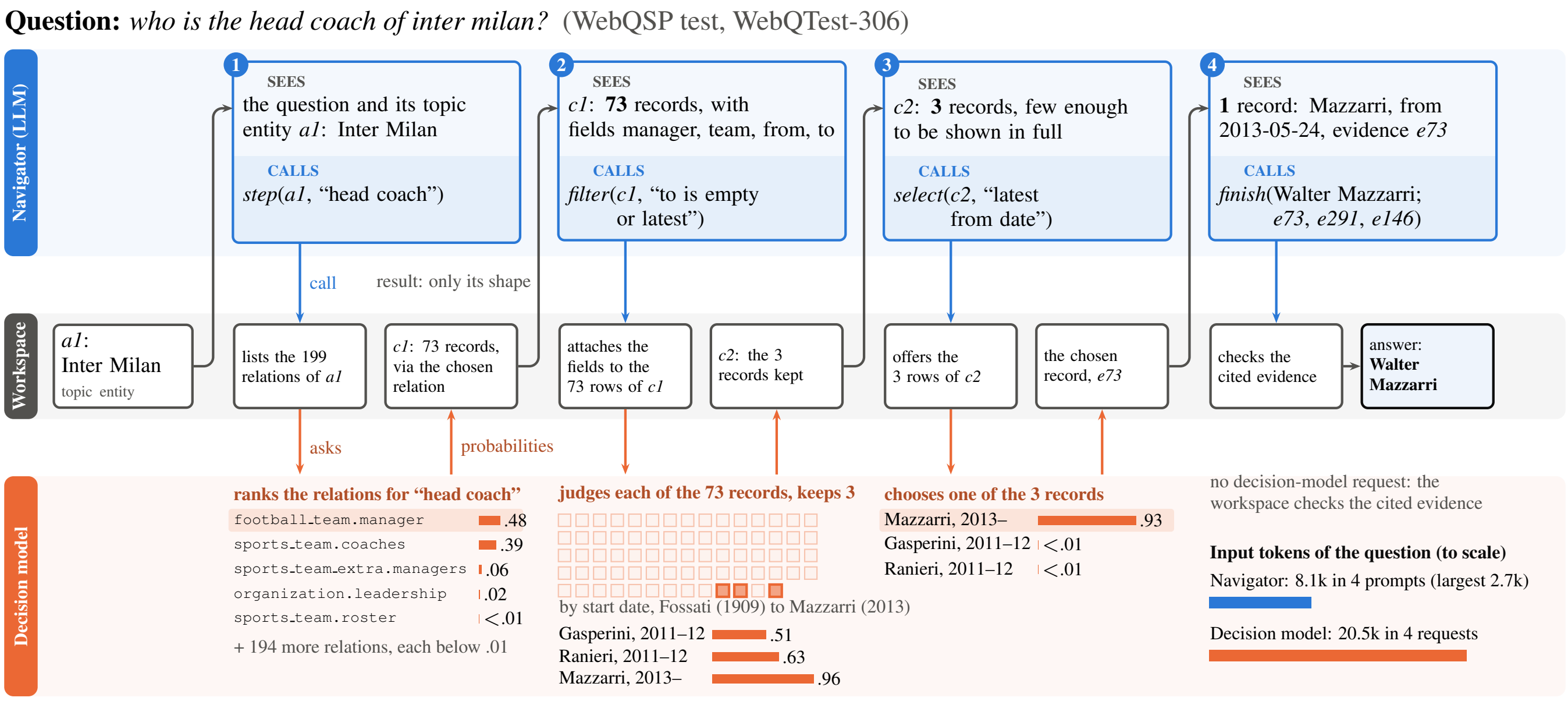

Knowledge graph question answering (KGQA) is the task of answering natural language questions by retrieving facts, entities, or values from a structured knowledge graph (KG) such as Freebase~\cite{bollacker2008freebase}.
Recent training-free methods use large language model (LLM) agents to explores the graph step by step~\cite{sun2024tog,jiang2023structgpt,sun2026sog}. At each step, the agent retrieves candidate relations, entities or triples, placing them in its context, and then decide how to continue.
The LLM thus does two jobs at once: it directs the search, and it reads everything the search returns. 
This kind of agentic approach reaches great answering accuracy, but at what cost?

As a matter of fact, most of this reading is wasted.
To explain this, let's consider the example in Figure~\ref{fig:overview}. A single entity can have hundreds of relations, such as the 199 of Inter Milan in Step 1 of Figure~\ref{fig:overview}, of which only a few fit the question. If an agent reads all relationships to select the most appropriate, its context would be filled within a few steps with hundreds of records, while the decision it requires is trivial.

Existing agents handle this in one of two ways.
Think-on-Graph (ToG) prunes relations per step and samples entities whenever twenty or more appear~\cite{sun2024tog}, which keeps the context small but can drop the answer before the LLM sees it.
Search-on-Graph (SoG) lets the LLM read the full neighborhoods~\cite{sun2026sog}, which makes it the most accurate training-free method, but its context grows with the graph, and long contexts cost tokens and money and are known to distract LLMs~\cite{liu2024lost,shi2023distracted}.

Yet most of this reading is a choice among given options, such as which relation fits the goal or whether a row meets a condition.
For this reason we propose to delegate these choices to a decision model, which reads an input and returns a probability for each option of a fixed choice~\cite{d3_2026}.
Since it generates no text, such a model makes these choices directly and at a lower cost per token.

Our agent therefore separates directing the search from reading its results, much as a rally crew splits the work between a navigator, who calls the route, and a driver, who makes every fine-grained choice on the road (Figure~\ref{fig:overview}).
Our \emph{navigator} is an LLM that directs the search.
It writes goals and conditions and calls four tools, \tool{step}, \tool{filter}, \tool{select} and \tool{finish}. The tools act on \emph{collections}, named sets of results kept outside of the navigator's context. After each call, the navigator gets back only a compact summary of the result.
The decision model is the \emph{driver} and does the reading behind the tools. It chooses the relations to follow among hundreds, judges each row of a collection against a condition, and selects the row that best meets a criterion.

This division of labour eases the trade-off between accuracy and cost.
On four Freebase benchmarks, our agent matches the accuracy of SoG, while its LLM reads 4 to 6 times fewer tokens.
Even adding the decision model's tokens, our agent reads fewer tokens than SoG, and most of them go to the decision model, which is much cheaper per token.
As a result, on paid APIs, a run of 1{,}000 WebQSP questions using our agent costs less than half as much as a run using SoG.

Our contributions are the following:
\begin{enumerate}
  \item A KGQA agent in which an LLM navigator directs the search and a decision model evaluates what the search retrieves, such as which relation to follow or which rows meet a condition (Section~\ref{sec:method}).
  \item A controlled comparison with ToG and SoG on four benchmarks, showing that our agent matches the accuracy of SoG at a fraction of its cost, also on entities with large neighborhoods (Sections~\ref{sec:results} to~\ref{sec:large}).
  \item Ablations of both models, showing that the decision model needs only 4B parameters, while the navigator needs about 27B (Sections~\ref{sec:ablation} to~\ref{sec:thinking}).
\end{enumerate}

\section{Related Work}
\label{sec:related}

\paragraph{Semantic parsing and subgraph retrieval.}
Earlier KGQA systems either translate the question into a logical form or SPARQL query that is executed on the graph~\cite{ye2022rngkbqa,yu2023decaf,luo2024chatkbqa}, or retrieve a question-specific subgraph and rank its entities~\cite{sun2018graftnet,he2021nsm,jiang2023unikgqa}; see~\cite{lan2021survey,gu2022kbqasurvey} for surveys.
Both are trained on the questions of each benchmark, and both fix what they look at before reasoning: a parser commits to a full query, and a retriever to a subgraph.
LLM agents instead explore the graph step by step and decide each step from what the previous one returned.

\paragraph{Training-free LLM agents.}
Think-on-Graph (ToG) lets a frozen LLM run a beam search, scoring candidate relations and entities in its prompt~\cite{sun2024tog}.
Later agents add planning and memory~\cite{chen2024pog}, generated facts~\cite{xu2024gog}, text retrieval~\cite{ma2025tog2}, tree search~\cite{song2025rekgmcts} or retrieved exemplars~\cite{xu2025egp}.
StructGPT~\cite{jiang2023structgpt}, Interactive-KBQA~\cite{xiong2024interactivekbqa} and Search-on-Graph (SoG)~\cite{sun2026sog} give the LLM tools over the graph, and SoG lets it choose every step from the full history and the current neighborhood.
All of them put what they retrieve into the LLM's prompt and differ in how they bound it: they show relation names before triples~\cite{jiang2023structgpt}, rank candidates with embeddings~\cite{xiong2024interactivekbqa}, sample or prune candidates~\cite{sun2024tog}, or, in SoG, return only the relation list for entities with more than 50 neighbors and truncate results at 1{,}000 rows~\cite{sun2026sog}.
EffiQA~\cite{dong2025effiqa} hands pruning to a second, fine-tuned model, but the LLM still reads the relations that remain.
Our agent is training-free as well, but its LLM reads no relation list and no large result.

\paragraph{Fine-tuned LLM agents.}
A second line moves the knowledge of the graph into the LLM's weights.
KG-Agent~\cite{jiang2025kgagent} fine-tunes a 7B LLM on synthesized tool-use programs, and keeps entity sets in variables that its tools act on, as our collections do.
GraphWalker~\cite{xu2026graphwalker} trains a 7B model on synthetic random-walk trajectories and then with reinforcement learning on CWQ, using tools that return the relations of an entity and the triples of chosen relations.
SoG-R1~\cite{sun2026sogr1} keeps the interface of SoG and trains an 8B model on each dataset, first on traces of a 235B teacher and then with reinforcement learning; above 50 rows its tool returns a paged list of relations.
These agents are the most accurate on WebQSP and CWQ (Table~\ref{tab:main}), but they need training questions and must be retrained for another graph or model, and the trained model still reads the relations and triples its tools return.
We train neither model, and change the navigator or the decision model by changing an API call (Sections~\ref{sec:ablation} and~\ref{sec:navigator}).

\paragraph{Separating planning from reading.}
General tool-using agents also read every observation in their prompt~\cite{yao2023react,schick2023toolformer}.
Since long and noisy prompts lower accuracy~\cite{liu2024lost,shi2023distracted,hsieh2024ruler}, they truncate tool outputs~\cite{yao2023react,qin2024toolllm} or redesign them: SWE-agent resolves more issues when its tools return short summaries~\cite{yang2024sweagent}.
These remedies, like those of the KGQA agents above, decide \emph{what} the LLM reads; we change \emph{who} reads.
Delegation itself is not new: LLMs call specialist models for subtasks~\cite{shen2023hugginggpt,suris2023vipergpt}, plan all steps before seeing any observation~\cite{xu2023rewoo}, or route whole queries to cheaper models~\cite{chen2024frugalgpt,ong2025routellm}.
Our agent differs from these in three ways.
First, the LLM still decides after every step, but sees only the shape of each result: its size, the chosen relation with two alternatives, and the rows only when there are few.
Second, a single decision model answers all reading questions, as choices among given options, instead of one module per subtask; it acts like a reranker that scores each candidate~\cite{nogueira2020monot5}, but also judges rows against a condition and selects among them.
Third, neither model is trained for KGQA.

\section{Preliminaries}
\label{sec:prelim}

\paragraph{Knowledge graph and records.}
A knowledge graph $\mathcal{G}$ is a set of facts $(s, r, o)$, where $s$ is an entity, $r$ a relation and $o$ an entity or a literal (a date, number or string).
Freebase stores many $n$-ary facts, such as a term in office or a film role, through unnamed mediator nodes (CVTs)~\cite{bordes2015simplequestions}.
We call these nodes \emph{records} and their outgoing relations \emph{fields}.
In Figure~\ref{fig:overview}, each of the 73 values of \tool{soccer.football\_team.manager} is a record with the fields manager, team, from and to, and the answer is a field value of one of them.
Because the Freebase dump does not mark mediator types, we treat an unnamed entity as a record when, for one of its types, fewer than half of 200 sampled instances have a name.

\paragraph{Task.}
Given a question $q$ and its topic entities $T$, the agent returns a set of answers $A$ and a set of evidence facts $F \subseteq \mathcal{G}$ such that every answer in $A$ occurs in a fact of $F$.
As in prior work~\cite{sun2024tog,sun2026sog}, we use gold topic entities, so entity linking is outside the scope of this paper.

\paragraph{Decision model.}
A decision model $\mathcal{D}$ reads a context $x$ and a finite set of options $O$ and returns a probability distribution $p = \mathcal{D}(x, O)$ over $O$, without generating text~\cite{d3_2026}.
Its decision is the most probable option, and the probabilities order the remaining options.
One request may contain several independent choice questions that share a context.

\section{Methods}
\label{sec:method}

The agent rests on one division of labour: the navigator decides \emph{what} to look for, and the decision model decides \emph{which} of the retrieved items fits.
The navigator is an LLM agent that plans the search, writes goals and conditions, and cites evidence.
It never sees the lists that it would otherwise have to read, such as the 199 relations of Inter Milan or its 73 manager records in Figure~\ref{fig:overview}; it sees only the \emph{shape} of each result.
The decision model reads these lists, but only to answer choice questions that the tools ask it.
Below, we describe the workspace that keeps results out of the navigator's prompt, the three operations through which the decision model reads, and the navigator, with its loop and the tools built on these operations, following the question of Figure~\ref{fig:overview} throughout.

\subsection{Workspace and Result Shapes}
\label{sec:workspace}

The workspace holds the state of one question outside the navigator's prompt.
It stores three kinds of objects, each with an identifier that the navigator uses in tool calls:
\emph{answer references} to an entity or a literal;
\emph{evidence references} to a retrieved fact;
and \emph{collections}, which are lists of rows.
A row holds an answer reference, the evidence reference of the fact that retrieved it, its \emph{origin} (the row of the previous collection from which it was reached), and any fields attached to it.
We call the entities in the rows of a collection its \emph{members}.
At the start, the workspace holds one answer reference per topic entity.

After every tool call, the navigator receives the shape of the result (Figure~\ref{fig:overview}, top):
the collection identifier and its number of rows;
the relation that was followed and the next two relations as alternatives;
for records, the names of the fields already attached and of those still available;
and a flag when the result was cut at the member budget.
A result with at most five rows is shown in full, with references, names and fields.
For a larger result, the navigator sees only the names of at most 20 named rows and the number of the others; records have no names and are not listed.
The navigator therefore never receives relation lists or the fields of large collections, and the size of what a tool returns to it is bounded by these constants rather than by the size of the result.

\subsection{Reading with the Decision Model: Choose, Judge, Reduce}
\label{sec:reading}

The decision model reads data through three operations only (Algorithm~\ref{alg:run}, bottom).
Every request contains the question and the navigator's goal or condition, and counts against the budget of decision-model requests (Section~\ref{sec:loop}).

\paragraph{Choose.}
Given a goal and options $O$, the decision model returns $p = \mathcal{D}(x, O)$, and the workspace orders the options by $p$.
Where stated below, $O$ contains a \emph{none} option, so that the decision model can reject all candidates.

\paragraph{Judge.}
Given a condition and a list of items, the decision model answers one independent question per item, with the options \emph{match} (the evidence supports the condition), \emph{no match} (the evidence contradicts it) and \emph{unknown} (the evidence is insufficient).
Up to 64 items share a request; a request larger than 60\,kB is split in half, and a single item that is too large is unknown.

\paragraph{Reduce.}
A list of more than 250 options, or one whose request would exceed 60\,kB, is split in half recursively into batches that fit.
The decision model ranks each batch, the workspace keeps the most probable 20\% of each batch (at least one option), and the procedure repeats on the survivors until a single batch remains, on which Choose makes the final decision.
Probabilities are compared only within a batch.
Unlike a cut to a fixed number of candidates before the model sees them, Reduce reads every option; it still prunes, since an option ranked low in its batch cannot return.

\subsection{The Navigator}
\label{sec:loop}

\begin{algorithm}[t]
  \caption{Navigator loop for one question.}
  \label{alg:navigator}
  \small
  \begin{algorithmic}[1]
    \REQUIRE question $q$, topic entities $T$, budgets $B$
    \STATE $W \gets$ workspace with answer references for $T$
    \STATE $M \gets [\,\text{system prompt}, (q, \text{references of } T)\,]$
    \WHILE{true}
      \STATE $f \gets$ (a budget in $B$ is spent)
      \IF{$f$}
        \STATE append a request to finish to $M$; allow only \tool{finish}
      \ENDIF
      \STATE $c \gets \text{LLM}(M)$ \hfill $\triangleright$ one tool call
      \STATE $o \gets \textsc{Run}(W, c)$ \hfill $\triangleright$ Algorithm~\ref{alg:run}
      \IF{$c$ is \tool{finish} and $o$ accepts it}
        \RETURN $(A, F)$ resolved by $W$ \hfill $\triangleright$ empty if abstained
      \ENDIF
      \IF{$f$}
        \RETURN $(\emptyset, \emptyset)$
      \ENDIF
      \STATE append $c$ without its reasoning text to $M$
      \STATE append the shape of $o$, or an error, to $M$ \hfill $\triangleright$ not $o$
    \ENDWHILE
  \end{algorithmic}
\end{algorithm}

The navigator is an LLM agent that answers a question by exploring the graph from its topic entities.
It works in a loop (Algorithm~\ref{alg:navigator}): at each turn, it reads the question and the history of its previous calls, reasons about what is still missing, and makes one tool call, for example to follow the relation from Inter Milan to its managers, to keep the managers whose term is current, or to answer.
The workspace runs the call, asks the decision model wherever the call needs a choice, and returns the shape of the result, from which the navigator plans its next call.
The navigator thus builds the answer step by step as a chain of collections, each derived from the previous one, without reading any of them in full.
From the size of a result, the relation that produced it and the fields it offers, it judges whether the search is on track; if not, it tries an alternative relation or another condition.
Its history keeps the system prompt, the question with the references of its topic entities, and each tool call with the shape of its result; the reasoning text of earlier turns is not kept.
The system prompt describes the tools and gives a few short rules, such as that a record is not an answer.
The loop ends when the navigator calls \tool{finish} with an answer that passes the check described below.
Budgets on navigator calls, decision-model requests and time bound the loop (Appendix~\ref{app:budgets}); when one is spent, the navigator gets one final turn in which it can only call \tool{finish}.

The navigator has four tools: \tool{step} follows a relation, \tool{filter} keeps the rows that meet a condition, \tool{select} picks the one row that best meets a criterion, and \tool{finish} returns the answer.
In each, the navigator states what it needs and the decision model makes the choice; \tool{finish} is checked in code.
Algorithm~\ref{alg:run} shows how the workspace runs each tool and what it asks the decision model; we describe the tools in turn.

\begin{algorithm}[t]
  \caption{How the workspace $W$ runs a tool call $c$ of the navigator for question $q$, and the operations through which the decision model $\mathcal{D}$ reads.}
  \label{alg:run}
  \small
  \textbf{function} \textsc{Run}$(W, c)$
  \begin{algorithmic}
    \IF{$c = \tool{step}(s, g)$}
      \IF{$g$ names a relation offered as an alternative}
        \STATE $r \gets g$ \hfill $\triangleright$ no request to $\mathcal{D}$
      \ELSE
        \STATE $O \gets$ relations of $s$ \hfill $\triangleright$ collection: of 3 members, with \emph{none}
        \STATE $r \gets \textsc{Choose}((q, g), O)$
        \IF{$r = \emph{none}$}
          \STATE $r \gets \textsc{Choose}((q, g), O \cup{}$ relations of 3 more members$)$
        \ENDIF
      \ENDIF
      \STATE retrieve $r$ for every member of $s$
      \RETURN entities as a new collection, or literals as a field of $s$
    \ELSIF{$c = \tool{filter}(C, k, \Phi)$}
      \STATE attach the fields $\Phi$ to the rows of $C$; apply the code checks
      \IF{the rows are plain entities}
        \STATE attach $\textsc{Choose}((q, k), \text{relations of } C)$ to every row
      \ENDIF
      \RETURN rows $x \in C$ with $\textsc{Judge}((q, k), C)[x] = \emph{match}$
    \ELSIF{$c = \tool{select}(C, k, \Phi)$}
      \STATE attach the fields $\Phi$ to the rows of $C$
      \RETURN $\textsc{Choose}((q, k), C \cup \{\emph{none}\})$
    \ELSIF{$c = \tool{finish}(A, E)$}
      \RETURN accept iff every answer in $A$ occurs in a fact of $E$
    \ENDIF
  \end{algorithmic}
  \bigskip
  \textbf{function} \textsc{Choose}$(x, O)$
  \begin{algorithmic}
    \WHILE{$|O| > 250$ or the request exceeds 60\,kB}
      \STATE split $O$ in half recursively into batches that fit \hfill $\triangleright$ Reduce
      \STATE $O \gets$ top 20\% of each batch $b$ by $\mathcal{D}(x, b)$
    \ENDWHILE
    \RETURN $\arg\max_{o \in O} \mathcal{D}(x, O)[o]$
  \end{algorithmic}
  \bigskip
  \textbf{function} \textsc{Judge}$(x, I)$
  \begin{algorithmic}
    \FOR{each item $i \in I$, in batches of at most 64}
      \STATE $\ell[i] \gets \arg\max \mathcal{D}((x, i), \{\emph{match}, \emph{no match}, \emph{unknown}\})$
    \ENDFOR
    \RETURN $\ell$
  \end{algorithmic}
\end{algorithm}

\paragraph{\tool{step}(source, goal).}
The source is an answer reference or a collection.
The workspace lists the relations of the source, each as an option with its name, direction, number of values, and whether the values are entities or literals, and the decision model chooses one for the goal with Choose.
For a collection, the options come from three evenly spaced members (first, middle, last) instead of all members, and include \emph{none}; if the decision model chooses \emph{none}, three further evenly spaced members are added and the decision model must choose from the merged list.
This keeps the request small but can miss a relation that only other members have.
The chosen relation is retrieved for every member.
Entity values form a new collection whose rows keep their origin; literal values are instead attached as a field to each member, so that a later \tool{select} or \tool{filter} returns the members themselves (the book with the earliest publication date rather than the date).
The two runner-up relations are returned to the navigator by name, without probabilities; if the chosen relation answers a different question, the navigator can repeat the step naming an alternative, or a field of the collection's records, and no decision-model request is made.
When the result contains records, the workspace reads the outgoing relations of three evenly spaced records, ignores schema relations, and reports them as available fields; a result of at most five records gets all fields attached.
In Figure~\ref{fig:overview}, the decision model chooses \tool{soccer.football\_team.manager} among the 199 relations of Inter Milan for the goal ``head coach'', and the navigator receives a collection of 73 records with the fields manager, team, from and to.

\paragraph{\tool{filter}(collection, condition, fields).}
The navigator states the condition and names the fields it needs, which the workspace attaches to every row, one bulk query per field.
Two optional checks in code then remove rows that cannot match, at no decision-model cost: a field must contain a given text (\emph{match}), or a record's period or date must overlap a given year or date (\emph{during}); a row whose values cannot be read is kept.
The decision model judges the rows with Judge.
If the rows are literals or carry fields, it judges them as they are.
If they are plain entities, the decision model first chooses a relation for checking the condition, as in \tool{step}; the workspace retrieves it for every member, and each member is judged on its values, a member without values being unknown.
The matching rows form a new collection.
Unknown rows are dropped and counted; when more than half of the rows are unknown, the navigator also receives the fields that could still be attached.
In Figure~\ref{fig:overview}, the navigator uses neither check, and the decision model judges all 73 records against the condition that the term has no end date or the latest one, keeping 3.

\paragraph{\tool{select}(collection, criterion, fields).}
With Choose, the decision model picks the one row that best meets a criterion such as ``the latest from date'', judged on the named fields, with \emph{none} among the options; Reduce applies to long collections.
The navigator receives the selected row in full, or an error with a short summary of the rows' contents if the decision model chooses \emph{none}.
The prompt tells the navigator to use \tool{select} only when the question asks for a single answer.
In Figure~\ref{fig:overview}, the decision model selects the record of Walter Mazzarri, who started in 2013, among the 3.

\paragraph{\tool{finish}(answer, evidence, text).}
The navigator answers with one answer reference and the evidence references that support it, or with a collection, in which case every row becomes an answer supported by the fact that retrieved it and the facts of its fields.
The workspace rejects an answer that does not occur in the cited evidence, a record given as an answer (and lists its field values instead), and a collection whose rows are mostly records.
A rejection returns an error, and the navigator may try again.
An answer without a reference is an abstention.

This check grounds the answer references: every entity or value in $A$ occurs in a retrieved fact of $F$.
The navigator also writes the answer as text in braces, which Hits@1 scores as in prior work (Section~\ref{sec:setup}); this text is not checked against the references.

\section{Results}
\label{sec:results}

We first describe the setup and report accuracy, then how much the navigator and the decision model read, how both change with the size of the neighborhoods the search explores, and what the agent costs on paid APIs.
We then vary the decision model, the navigator and its thinking, and analyse accuracy by question type and where the agent fails.

\subsection{Experimental Setup}
\label{sec:setup}

\paragraph{Datasets.}
Following ToG~\cite{sun2024tog} and SoG~\cite{sun2026sog}, we evaluate on four Freebase-based benchmarks that span a wide range of reasoning depth: from single-fact questions in SimpleQuestions~\cite{bordes2015simplequestions}, through the one- and two-hop questions of WebQSP~\cite{yih2016webqsp}, to the compositional questions with constraints and comparisons of CWQ~\cite{talmor2018cwq}, while GrailQA~\cite{gu2021grailqa} additionally tests generalization to unseen schema items.
We use the full test sets of WebQSP (1{,}639 questions) and CWQ (3{,}531), and the 1{,}000-question samples of GrailQA and SimpleQuestions drawn by ToG.
No question is filtered out: topic entities are given, and every query runs against the complete Freebase dump (Appendix~\ref{app:details}).

\paragraph{Models.}
Our navigator is GLM-5.3-Flash\footnote{\url{https://huggingface.co/zai-org/GLM-5.3-Flash}}, a mixture-of-experts LLM with 18B active parameters, run at temperature 0 with low reasoning effort.
The decision model is \dthree{}~\cite{d3_2026}, a 27B model used off the shelf, without fine-tuning.
Both roles are interchangeable: Section~\ref{sec:ablation} swaps the decision model for smaller models of its family and for Jev-1.14\footnote{\url{https://openrouter.ai/typesafe/jev-1.14}}, and Section~\ref{sec:navigator} swaps the navigator for five LLMs from 2B to 763B parameters.

\paragraph{Metrics.}
Following prior work~\cite{sun2024tog,sun2026sog}, we report Hits@1, scored with ToG's answer matching (Appendix~\ref{app:details}).

\paragraph{Baselines.}
We compare against published results of training-free and fine-tuned methods, although these use other LLMs and, in part, other question samples.
For a controlled comparison, we also run ToG from its released code and SoG from our reimplementation, as no code is public, giving both the same navigator LLM, graph and questions as our agent (Appendix~\ref{app:details}).

\subsection{Accuracy}

\begin{table}[t]
  \centering
  \small
  \setlength{\tabcolsep}{3pt}
  \resizebox{\columnwidth}{!}{%
  \begin{tabular}{@{}llcccc@{}}
    \toprule
    Method & LLM & WebQSP & CWQ & GrailQA & SQ \\
    \midrule
    \multicolumn{6}{@{}l}{\emph{Fine-tuned}} \\
    DecAF~\cite{yu2023decaf}        & -- & 82.1 & 70.4 & -- & -- \\
    ChatKBQA~\cite{luo2024chatkbqa} & Llama-2 & 83.2 & \underline{82.7} & -- & -- \\
    GraphWalker~\cite{xu2026graphwalker} & Qwen2.5-7B & \underline{91.5} & 79.6 & \underline{86.3} & -- \\
    SoG-R1~\cite{sun2026sogr1}      & Llama-3.1-8B & \textbf{91.7} & \textbf{88.7} & \textbf{90.0} & -- \\
    \midrule
    \multicolumn{6}{@{}l}{\emph{Training-free, published}} \\
    ToG~\cite{sun2024tog}           & GPT-3.5 & 76.2 & 57.1 & 68.7 & 53.6 \\
    ToG                             & GPT-4   & 82.6 & 67.6 & 81.4 & 66.7 \\
    ToG (by SoG)                    & GPT-4o  & 80.2 & 48.9 & 65.5 & 57.2 \\
    PoG~\cite{chen2024pog}          & GPT-4   & 87.3 & 75.0 & \underline{84.7} & -- \\
    REKG-MCTS~\cite{song2025rekgmcts} & GPT-4 & 86.0 & \underline{76.3} & -- & -- \\
    PoG-EGP~\cite{xu2025egp}        & Gemini-2.5-F & 88.6 & 75.4 & -- & -- \\
    Residual-guided~\cite{sei2026residual} & Qwen3-8B & 86.0 & 73.0 & -- & -- \\
    SoG~\cite{sun2026sog}           & GPT-4o  & \textbf{91.3} & 75.1 & \textbf{86.9} & \underline{84.8} \\
    SoG                             & Qwen3-235B & \underline{89.3} & \textbf{77.1} & 83.9 & \textbf{86.4} \\
    \midrule
    \multicolumn{6}{@{}l}{\emph{Training-free, our LLM, graph and questions}} \\
    ToG (our run)  & GLM-5.3-F & 71.8 & 54.7 & 66.7 & 49.7 \\
    SoG (our reimpl.) & GLM-5.3-F & \textbf{90.4} & \underline{79.4} & \underline{86.2} & \textbf{73.9} \\
    Ours           & GLM-5.3-F & \underline{89.7} & \textbf{81.6} & \textbf{87.4} & \textbf{73.9} \\
    \bottomrule
  \end{tabular}}
  \caption{Hits@1 (\%) on the test sets of WebQSP and CWQ and on ToG's samples of GrailQA and SimpleQuestions (SQ). Valid Hits@1 on SQ is in Appendix~\ref{app:sq}. Published results are copied from the cited papers; they use other LLMs and, in part, other test samples and answer-matching rules, and DecAF and ChatKBQA link entities themselves. GraphWalker and SoG-R1 fine-tune their LLM with supervised and reinforcement learning on training questions; SoG-R1 trains one model per dataset. SoG (our reimpl.): our implementation of SoG (Appendix~\ref{app:details}). Our runs use GLM-5.3-Flash at low reasoning effort. Within each group, bold marks the best result per benchmark and underline the second best. GLM-5.3-F: GLM-5.3-Flash.}
  \label{tab:main}
\end{table}

We first compare the three training-free methods run with our LLM, graph and questions (Table~\ref{tab:main}, bottom group), where differences can only come from the method.

Our method matches SoG, the strongest of the three, although its LLM reads none of the relation lists and neighbor tables that SoG places in its prompt.
On WebQSP, GrailQA and SimpleQuestions the two methods are within 1.2 points of each other and no difference is significant (exact McNemar test, $p \geq 0.21$); on CWQ, whose questions need the most steps, our method is 2.2 points higher ($p < 0.001$).
Delegating the reading thus costs no accuracy: of 50 inspected wrong answers on CWQ, only 4 come from the decision model's filtering and selection, and at most 6 more from a wrong relation (Appendix~\ref{app:errors}).
ToG instead keeps its prompt small by keeping at most three relations and entities per step, and it is 18 to 27 points below both methods.
What each method spends on this accuracy is the subject of Section~\ref{sec:cost}.

The published results use other LLMs, graph versions and in part other test samples, so they compare the methods only loosely.
Our reimplementation of SoG comes close to its published results on WebQSP and GrailQA, which suggests it is faithful.
Against the published training-free methods, ours is competitive with the best on WebQSP and GrailQA and above all of them on CWQ.
On SimpleQuestions it is below the published SoG, but our reimplementation of SoG shows the same gap, which therefore comes from the setup rather than the method (Appendix~\ref{app:sq}).

Fine-tuned methods remain stronger where they have training data: SoG-R1, which trains one 8B model per dataset with supervised and reinforcement learning, leads on all three benchmarks it reports, by up to 7.1 points on CWQ.
Our method fine-tunes neither of its models, and the same two models answer all four benchmarks.

\subsection{Cost}
\label{sec:cost}

\begin{table}[t]
  \centering
  \small
  \setlength{\tabcolsep}{2pt}
  \begin{tabular}{@{}lrrrr@{}}
    \toprule
     & WebQSP & CWQ & GrailQA & SimpleQ. \\
    \midrule
    \multicolumn{5}{@{}l}{\textbf{Our method}} \\
    \multicolumn{5}{@{}l}{\emph{Navigator (GLM-5.3-Flash)}} \\
    LLM Calls                         & \textbf{3.0}  & \textbf{5.2}  & \textbf{3.2}  & \textbf{2.4} \\
    Input tokens, total (k)       & \textbf{5.5}  & \underline{10.6} & \underline{5.9}  & \underline{3.9} \\
    Output tokens (k)             & \textbf{0.13} & \textbf{0.26} & \textbf{0.15} & \textbf{0.10} \\
    \addlinespace[3pt]
    \multicolumn{5}{@{}l}{\emph{Decision model (\dthree{})}} \\
    Requests                      & 2.6  & 5.4  & 2.1  & 1.8 \\
    Input tokens (k)              & 13.7 & 25.1 & 7.8  & 7.1 \\
    Share of all input tokens (\%)& 69.9 & 66.2 & 53.3 & 63.1 \\
    \midrule[\heavyrulewidth]
    \multicolumn{5}{@{}l}{\textbf{Baselines, same LLM}} \\
    \multicolumn{5}{@{}l}{\emph{ToG (our run)}} \\
    LLM calls                     & 5.8  & \underline{8.7}  & 4.5  & 3.9 \\
    Input tokens, total (k)       & \underline{6.1}  & \textbf{8.6}  & \textbf{3.6}  & \textbf{2.7} \\
    Output tokens (k)             & 0.69 & 1.19 & 0.55 & 0.43 \\
    \addlinespace[3pt]
    \multicolumn{5}{@{}l}{\emph{SoG (our reimpl.)}} \\
    LLM calls                     & \underline{5.5}  & 10.0 & \underline{4.2}  & \underline{3.1} \\
    Input tokens, total (k)       & 25.8 & 64.3 & 25.3 & 20.5 \\
    Output tokens (k)             & \underline{0.25} & \underline{0.51} & \underline{0.19} & \underline{0.13} \\
    \bottomrule
  \end{tabular}
  \caption{Cost per question, as means. All our runs use the navigator at low reasoning effort. Input tokens, total: the input of all LLM calls of a question, summed. Output tokens include reasoning tokens. Navigator tokens are as reported by the provider. Decision-model input tokens: the state and questions of each request, counted once per request. Bold: lowest LLM cost per column among the three methods (navigator for ours); underline: second lowest.}
  \label{tab:cost}
\end{table}

\begin{table}[t]
  \centering
  \small
  \setlength{\tabcolsep}{3pt}
  \scalebox{1.1}{%
  \begin{tabular}{@{}lrrrrr@{}}
    \toprule
     & & \multicolumn{2}{c}{LLM (k)} & DM & \$ per \\
    \cmidrule(lr){3-4}
     & Hits@1 & in & out & in (k) & 1k q. \\
    \midrule
    Ours (Jev-1.14) & 89.7 & 6.2  & 0.14 & 20.3 & \textbf{1.85} \\
    SoG (reimpl.)   & 90.4 & 25.8 & 0.25 & --   & 3.99 \\
    \bottomrule
  \end{tabular}}
  \caption{API cost on WebQSP, both with GLM-5.3-Flash at low reasoning effort, at OpenRouter list prices of 17 November 2026: GLM-5.3-Flash \$0.15 and \$0.50 per 1M input and output tokens, Jev-1.14 \$0.042 per 1M input tokens, output not billed. Tokens are means per question as reported by the APIs; Jev's count uses its own tokenizer and is therefore higher than the counts of Table~\ref{tab:cost}. Since tokens are in thousands per question, 1k questions use them in millions; for example, our cost per 1k questions is $6.2 \times \$0.15 + 0.14 \times \$0.50 = \$1.00$ for the LLM plus $20.3 \times \$0.042 = \$0.85$ for Jev, \$1.85 in total, and SoG's is $25.8 \times \$0.15 + 0.25 \times \$0.50 \approx \$3.99$ (computed from unrounded means). No prompt-cache discount is applied. DM: decision model.}
  \label{tab:api_cost}
\end{table}

In our method, the decision model reads most of the input, 53 to 70\% of all input tokens (Table~\ref{tab:cost}), while the navigator reads 3.9k to 10.6k tokens per question, of which about 1.5k per call are the fixed system prompt and tool descriptions.
The two loads behave differently.
The decision model's input spans two orders of magnitude, from 0.6k to 76k tokens per question between the 5th and 95th percentiles, because a relation list has a median of 168 to 178 entries and a filter can judge hundreds of rows.
The navigator's input does not follow this variation, because a tool returns the size of a collection and at most a few of its rows, not its content.

SoG puts all of this reading into the LLM, which reads 4.3 to 6.1 times as many input tokens as our navigator, in line with its published runs with GPT-4o (15.0k to 54.6k tokens).
Even counting the decision model, our method reads fewer tokens than SoG (11.0k to 35.7k per question), and most of them go to a smaller model that generates no text.
ToG's LLM reads about as few input tokens as our navigator, but it gets there by pruning: it generates 3.7 to 5.3 times more output tokens and answers 18 to 27 points fewer questions correctly.
The three methods thus make different trade-offs: ToG keeps the LLM's input small by reading less, SoG reaches high accuracy by letting the LLM read everything, and our method keeps both the accuracy and the small input by letting another model read.

Token counts do not price the two models alike, so we also price a whole run on paid APIs (Table~\ref{tab:api_cost}).
For this we replace \dthree{} by Jev-1.14, a decision model served through a paid API, which reaches the same Hits@1 on WebQSP (Section~\ref{sec:ablation}).
At OpenRouter list prices, 1{,}000 WebQSP questions cost \$1.85 with our method and \$3.99 with SoG, 2.2 times as much, at the same accuracy ($p = 0.36$).
The difference comes from SoG's prompt: our method moves most of the reading to a model whose input costs 3.6 times less per token and whose output is not billed.
With the self-hosted \dthree{} of the other runs, counting only the LLM, SoG costs 4.0 to 5.8 times as much as our method on the four benchmarks.

\subsection{Large Neighborhoods}
\label{sec:large}

\begin{figure*}[t]
  \centering
  \includegraphics[width=\textwidth]{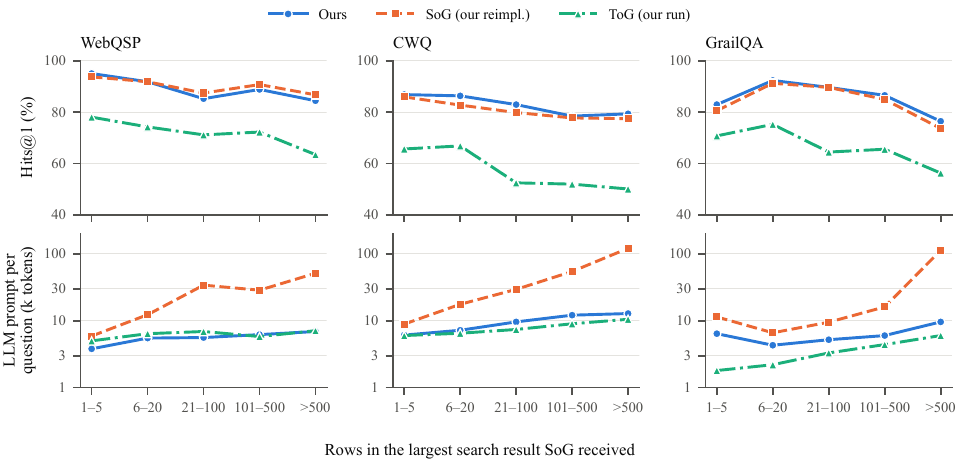}
  \caption{ToG, SoG and our agent, all with GLM-5.3-Flash, on questions grouped by the size of the largest search result SoG received (rows before truncation), a measure of the neighborhoods on the question's path that does not depend on our agent's search. Top: Hits@1. Bottom: mean LLM prompt tokens per question (log scale); for our agent, the navigator only. Groups contain 41 to 1{,}288 questions. SimpleQuestions is left out, because there the size of a result mostly reflects how many objects answer the question (Appendix~\ref{app:sq}).}
  \label{fig:scale_baselines}
\end{figure*}

Large neighborhoods are where reading costs the most, so they test whether delegating it holds up.
To compare the methods on the same questions without grouping them by our own search, we group the questions by the largest search result SoG received, i.e.\ the largest neighborhood on the path SoG took (Figure~\ref{fig:scale_baselines}).

Large neighborhoods make questions harder for every method.
From the smallest to the largest group, the Hits@1 of SoG and of our method falls by 6 to 11 points, about in parallel, and on the questions above 100 rows neither method is significantly better ($p \geq 0.08$).
What differs is the cost of this accuracy.
From the smallest to the largest group, SoG's prompt per question grows 9 to 14 times, to 51k to 121k tokens, while the navigator's grows at most 2.1 times, so on the largest neighborhoods SoG's LLM reads 7 to 12 times as many tokens as our navigator.
In our method the extra reading goes to the decision model.
ToG keeps its prompt small by keeping at most three relations per step, and it is 10 to 31 points below both methods in every group.
Large collections built by our own search are a different matter, which we discuss in Section~\ref{sec:limitations}.

\subsection{Reading Needs Only a Small Model}
\label{sec:ablation}

\begin{table}[t]
  \centering
  \small
  \setlength{\tabcolsep}{3pt}
  \begin{tabular}{@{}lrccc@{}}
    \toprule
    Decision model & Params & Hits@1 & Navigator calls & DM req. \\
    \midrule
    d3-edge  & 0.6B & 77.8 & 5.4 & 3.9 \\
    d3-lite  & 0.8B & 81.6 & 4.8 & 3.5 \\
    d3-nano  & 2B   & 84.3 & 4.3 & 3.2 \\
    d3-mini  & 4B   & 88.7 & 3.4 & 2.9 \\
    d3-flash & 9B   & 89.4 & \underline{3.1} & \underline{2.7} \\
    d3-turbo & 27B  & \textbf{90.1} & \textbf{3.0} & \textbf{2.6} \\
    \dthree{} (Table~\ref{tab:main}) & 27B & \underline{89.7} & \textbf{3.0} & \textbf{2.6} \\
    \midrule
    Jev-1.14 & n/a & \underline{89.7} & 3.2 & \underline{2.7} \\
    \bottomrule
  \end{tabular}
  \caption{WebQSP test results (\%) with \dthree{} replaced by other decision models: the other models of its family, and Jev-1.14, used through its API, whose size is not published. The navigator (GLM-5.3-Flash), budgets and protocol are unchanged. Calls and requests are means per question. Bold: best per column; underline: second best (higher Hits@1, fewer calls and requests). Tied values are marked alike. DM: decision model.}
  \label{tab:ablation}
\end{table}

The decision model does the reading, so its quality bounds what the navigator can find.
To measure how much capacity the reading needs, we replace \dthree{} by the other decision models of its family~\cite{d3_2026} and run WebQSP with everything else unchanged (Table~\ref{tab:ablation}).
The reading needs only a small model: from 4B parameters upward, including the newer 27B d3-turbo, the decision models are within one point of \dthree{} ($p \geq 0.13$), while those of 2B and fewer parameters lose 5.4 to 11.9 points ($p < 0.001$).
A weaker reader also makes the navigator work more: the smaller the decision model, the more calls the navigator makes and the more requests the decision model receives, and with the 0.6B model the agent gives no answer on 7.1\% of the questions instead of 0.7\%.
Reading errors thus cost twice, in accuracy and in navigator calls.
The agent does not depend on \dthree{} either: Jev-1.14 reaches the same Hits@1 ($p = 1.0$) with about the same number of calls.

\subsection{Navigating Needs a Larger Model}
\label{sec:navigator}

\begin{table}[t]
  \centering
  \small
  \setlength{\tabcolsep}{2.5pt}
  \begin{tabular}{@{}lrcccc@{}}
    \toprule
    Navigator & Params & Hits@1 & Calls & Input & No answer \\
    \midrule
    GLM-5.3-Flash        & 320B & \textbf{88.7} & 3.4  & \underline{6.5}  & \textbf{1.3} \\
    GLM-5.3              & 753B & 87.6 & \textbf{3.2}  & \textbf{5.8}  & 1.8 \\
    DeepSeek-V4.1-Flash & 763B & \underline{87.9} & 3.7  & 8.2  & \underline{1.6} \\
    Qwen3.8-27B          & 27B  & 87.1 & \underline{3.3}  & 7.4  & 1.9 \\
    Qwen3.5-9B           & 9B   & 84.6 & 4.7  & 10.9 & 2.8 \\
    Qwen3.5-2B           & 2B   & 70.0 & 10.1 & 26.9 & 10.5 \\
    \bottomrule
  \end{tabular}
  \caption{WebQSP test results (\%) with other navigators and d3-mini (4B) as the decision model; the GLM-5.3-Flash row is the d3-mini run of Table~\ref{tab:ablation}. Calls and the total navigator input tokens (thousands) are means per question; No answer is the share of questions without an answer (abstained, or no answer when a budget ran out). Params are total parameters. The Qwen3.5 navigators run with thinking on (Section~\ref{sec:thinking}). Bold: best per column; underline: second best (higher Hits@1; fewer calls, input tokens and questions without an answer).}
  \label{tab:navigator}
\end{table}

Directing the search is the LLM's job, so we next measure how much capacity it needs.
We replace GLM-5.3-Flash by five LLMs from 2B to 763B parameters and keep d3-mini, the smallest decision model within one point of \dthree{}, with everything else unchanged (Table~\ref{tab:navigator}).
Navigating needs a larger model than reading: navigators of 27B parameters and more are within 1.6 points of GLM-5.3-Flash, and only Qwen3.8-27B is significantly lower ($p = 0.016$), while the 9B navigator loses 4.1 points and the 2B navigator 18.7 ($p < 0.001$).
A larger LLM brings nothing: GLM-5.3, more than twice the size of GLM-5.3-Flash, does not improve on it.
Since the navigator reads no large results, the smaller navigators lose their way more often: they make up to three times as many calls and more often end without an answer.
Reading thus works from 4B parameters while navigating needs about 27B, and the split puts the larger model only where it is needed.
The whole agent can therefore run on two small open models: with Qwen3.8-27B as the navigator and d3-mini (4B) as the decision model, it reaches 87.1\% on WebQSP, 2.6 points below our main configuration and 3.3 points below SoG with a 320B LLM.
Both models together can be self-hosted on a single GPU server, and since smaller models need less computation per token, this configuration should also lower the energy use and emissions per question, although we did not measure them.

\subsection{Thinking Helps Small Navigators, but Does Not Replace Size}
\label{sec:thinking}

\begin{table}[t]
  \centering
  \small
  \setlength{\tabcolsep}{2pt}
  \begin{tabular}{@{}llccccc@{}}
    \toprule
    Navigator & Thinking & Hits@1 & Calls & Input & Output & No answer \\
    \midrule
    \multirow{2}{*}{Qwen3.5-9B} & off & \underline{79.0} & \underline{5.5}  & \underline{14.9} & \textbf{0.4} & \underline{8.3} \\
                                & on  & \textbf{84.6} & \textbf{4.7}  & \textbf{10.9} & \underline{1.2} & \textbf{2.8} \\
    \midrule
    \multirow{2}{*}{Qwen3.5-2B} & off & 57.4 & 11.0 & 35.2 & 2.0 & 28.7 \\
                                & on  & 70.0 & 10.1 & 26.9 & 3.6 & 10.5 \\
    \bottomrule
  \end{tabular}
  \caption{WebQSP test results (\%) of the Qwen3.5 navigators with thinking off (greedy decoding) and on (the sampling recommended for thinking), with d3-mini as the decision model. Input and output are navigator tokens per question in thousands; No answer is the share of questions without an answer; the other columns are as in Table~\ref{tab:navigator}. Bold: best per column; underline: second best (higher Hits@1; lower values in the other columns).}
  \label{tab:thinking}
\end{table}

Since small LLMs fail at planning the search, we test whether thinking helps them plan, running the two Qwen3.5 navigators also with thinking off (Table~\ref{tab:thinking}).
Thinking raises Hits@1 by 5.6 points for the 9B navigator and by 12.6 points for the 2B navigator ($p < 0.001$), mainly because they less often end without an answer.
Even with thinking, both stay below the navigators of 27B parameters and more: thinking narrows the gap but does not replace capacity.

\subsection{Question Types}

\begin{table}[t]
  \centering
  \small
  \setlength{\tabcolsep}{4pt}
  \begin{tabular}{@{}llrccc@{}}
    \toprule
    Dataset & Type & $n$ & ToG & SoG & Ours \\
    \midrule
    \multirow{4}{*}{CWQ} & Composition  & 1{,}546 & 55.6 & \underline{79.6} & \textbf{82.0} \\
                         & Conjunction  & 1{,}575 & 56.1 & \underline{81.1} & \textbf{82.1} \\
                         & Comparative  & 213     & 59.2 & \underline{80.3} & \textbf{87.3} \\
                         & Superlative  & 197     & 33.0 & \underline{63.5} & \textbf{67.5} \\
    \midrule
    \multirow{3}{*}{GrailQA} & I.i.d.        & 240 & 62.5 & \underline{85.0} & \textbf{86.7} \\
                             & Compositional & 198 & 51.0 & \underline{75.3} & \textbf{79.3} \\
                             & Zero-shot     & 562 & 74.0 & \textbf{90.6}    & \textbf{90.6} \\
    \bottomrule
  \end{tabular}
  \caption{Hits@1 (\%) by question type. ToG (our run) and SoG (our reimpl.) use the same LLM, graph and questions as our agent. Bold: best per row; underline: second best; tied values are marked alike.}
  \label{tab:types}
\end{table}

By question type (Table~\ref{tab:types}), our method is above SoG on every CWQ type, most on comparative and superlative questions, which compare or rank candidates by a value and which our navigator delegates to the decision model with \tool{filter} and \tool{select}, although these groups are small.
Superlative and compositional questions remain the hardest for all three methods, and on GrailQA zero-shot questions are not harder than i.i.d.\ ones for either training-free method.

\section{Discussion}
\label{sec:discussion}

\paragraph{Reading and directing are separable.}
Our results suggest that a KGQA agent does two jobs that need different models.
Reading, i.e.\ choosing relations and judging rows, is a narrow decision over given candidates, and a 4B decision model does it as well as a 27B one.
Directing the search, i.e.\ deciding which step to take next and when to stop, needs an LLM of about 27B parameters, and thinking does not make up for a smaller one.
A method that puts both jobs into one LLM must size it for the harder job and then pays its price for the easier one as well.
Splitting them lets each job run on the smallest model that does it well.

\paragraph{Delegate or pay.}
The three training-free methods trade accuracy against LLM input in different ways.
ToG keeps the prompt small by pruning and loses accuracy; SoG keeps the accuracy by letting the LLM read everything and pays with a prompt that grows with the neighborhood; our agent keeps both by moving the reading to a model that is cheaper per token and generates no text.
The saving thus comes not from reading less but from reading elsewhere, and it is largest where reading costs most: on large neighborhoods the navigator's prompt barely grows while SoG's grows about tenfold.
Since the decision model's cost does not depend on the navigator, the saving would grow further with a stronger and more expensive LLM as the navigator.

\paragraph{The interface becomes the bottleneck.}
Once the navigator no longer reads, its errors come from planning: in our sample of wrong answers, the decision model causes few, and the agent's other errors are wrong relations, early stops and loops.
Several of them trace back to the interface, such as the missing tool to intersect two collections, and the largest failures arise when the agent's own search builds collections too large for its request budget (Section~\ref{sec:limitations}).
What the navigator can do is thus bounded by the tools it is given, and improving them looks like a cheaper lever than a larger LLM, which brought no gain in our experiments.

\paragraph{Training and other sources.}
Fine-tuned methods remain more accurate where training data exist.
The split suggests where training would pay off: the reading is already handled by a general decision model, so training could target the navigator alone, whose job is smaller than answering from the raw graph.
More generally, the same division applies wherever an agent searches structured data larger than its context, such as tables, relational databases or APIs, with an LLM that plans over the shape of results and a smaller model that reads them.

\section{Limitations}
\label{sec:limitations}

We use gold topic entities, as ToG and SoG do, so the results do not include entity linking.
Our main results use one navigator LLM at low reasoning effort; we vary the decision model and the navigator only on WebQSP, run the two smallest navigators, which have no effort levels, with thinking on and off under different decoding, and have not measured how the results change with a navigator that reads the raw results itself.
Some published baselines use other question samples or graph versions, so only the rows marked as run by us are a controlled comparison.
Our SoG is a reimplementation with our own few-shot examples, since its code and examples are not published.
Our agent fails most often when its own search builds large collections: on CWQ, Hits@1 falls from 85\% when every collection has at most five rows to 42\% when one has more than 500 rows, and on these questions, which partly reflect the navigator's own wrong steps, it is 17.1 points below SoG ($p < 0.001$).
The cause is the scale of the reading rather than its quality: filtering 2{,}000 records at 64 rows per request needs more decision-model requests than a question may use, a single request over such a collection can time out, and the navigator may repeat a step until a budget runs out; larger request budgets, batched requests or a tool that intersects collections could address this.
Finally, the per-question budgets were set during development on training and validation questions and were not tuned per dataset.

\section{Conclusion}
\label{sec:conclusion}

This paper asked whether an LLM agent that navigates a knowledge graph must also read the hundreds of relations and records that its search retrieves, and our results suggest that it need not.
Most of this reading serves to choose among given options, such as which relation to follow or which records meet a condition, and a decision model can make such choices at a fraction of the cost.
In our agent, an LLM navigator directs the search through four tools and sees only a compact summary of each result, while a decision model behind the tools evaluates what the search retrieves.
On four Freebase benchmarks, our agent matches the accuracy of SoG while its LLM reads 4 to 6 times fewer tokens.
Even adding the decision model's tokens, the agent reads fewer tokens than SoG, and a run of 1{,}000 WebQSP questions on paid APIs costs less than half as much.
The ablations also show that the decision model needs only 4B parameters, while the navigator needs about 27B.
Future work will give the navigator tools for the large collections that its own search can build, and bring the same division of labour to other structured sources, such as tables and relational databases.

\bibliographystyle{plain}
\bibliography{references}

\appendix

\section{Budgets and Interface Constants}
\label{app:budgets}

\begin{table}[h]
  \centering
  \small
  \setlength{\tabcolsep}{2pt}
  \begin{tabular}{@{}lr@{\hspace{6pt}}lr@{}}
    \toprule
    \multicolumn{2}{@{}l}{\emph{Budgets per question}} & \multicolumn{2}{l@{}}{\emph{Interface}} \\
    \midrule
    Navigator calls            & 30      & Rows shown in full          & 5 \\
    Navigator tokens (input)   & 60k    & Row names shown             & 20 \\
    DM requests             & 25      & Alternatives returned       & 2 \\
    DM request size         & 60\,kB  & Options per Choose request  & 250 \\
    Graph queries           & 2{,}000 & Items per Judge request     & 64 \\
    Members per step        & 2{,}000 & Kept per Reduce batch       & 20\% \\
    Retrieved facts         & 20{,}000& Sampled members, records    & 3 \\
    Time                    & 10 min  &                             &   \\
    \bottomrule
  \end{tabular}
  \caption{Budgets and interface constants, fixed for all datasets. DM: decision model.}
  \label{tab:budgets}
\end{table}

We set the budgets of the navigator loop and the constants of the interface (Table~\ref{tab:budgets}) during development on training and validation questions and use the same values for all datasets.
The final turn of Section~\ref{sec:loop} also starts 30 seconds before the time limit.
The budgets interact: filtering a collection of 2{,}000 records takes 32 Judge requests, more than the 25 decision-model requests allowed, which is one source of the errors on large collections discussed in Section~\ref{sec:limitations}.
The navigator is called with native function calling, one tool call per turn.
Identical tool calls are answered from a cache, and a malformed call returns an error message to the navigator at the cost of one navigator call.
The system prompt gives its rules without examples, for example that a filter on records for a question that names a year should check that year with \emph{during}.

\section{Experimental Details}
\label{app:details}

\paragraph{Datasets.}
CWQ builds its questions from WebQSP by composition, conjunction, comparison and superlatives.
ToG's GrailQA sample comes from the development set, since the test answers are hidden, and contains 562 zero-shot, 240 i.i.d.\ and 198 compositional questions.
Topic entities are the gold entities of the ToG release for CWQ, GrailQA and SimpleQuestions, and for WebQSP the topic entity of the first complete parse of each question.

\paragraph{Graph.}
We query the full Freebase dump prepared by the Freebase-Setup project\footnote{\url{https://github.com/dki-lab/Freebase-Setup}} in Virtuoso~7.2.5, read-only through SPARQL.
ToG's setup loads the same Freebase release into the same Virtuoso version but removes non-English triples; our graph keeps them.

\paragraph{Navigator.}
GLM-5.3-Flash belongs to the GLM-5 series~\cite{glm2026glm5} and has 320B total and 18B active parameters.
It always generates reasoning tokens before its answer, which we count as output tokens.
Its API offers three reasoning efforts (low, high and max) and uses max by default.
We use low effort for all runs, ours and both baselines'; at the default effort, many calls take minutes, which makes full runs of ToG impractical.

\paragraph{Decision models.}
\dthree{} is built on Qwen3.8-27B and returns a probability for every option of a choice question without generating text; we serve it with vLLM on one NVIDIA B200 GPU.
The ablation of Section~\ref{sec:ablation} uses the five smaller models of the family (0.6B to 9B) and the newer 27B model d3-turbo, all without fine-tuning; d3-turbo runs in its reference implementation in Transformers, since vLLM does not support its attention, and the others run in vLLM.
Jev-1.14, whose size is not published, is called through OpenRouter's decisions API.

\paragraph{Other navigators.}
The navigator ablation of Section~\ref{sec:navigator} keeps d3-mini (4B) as the decision model and uses GLM-5.3\footnote{\url{https://huggingface.co/zai-org/GLM-5.3}} (753B parameters), DeepSeek-V4.1-Flash\footnote{\url{https://huggingface.co/deepseek-ai/DeepSeek-V4.1-Flash-1017}} (763B), the dense Qwen3.8-27B\footnote{\url{https://huggingface.co/Qwen/Qwen3.8-27B}}, and Qwen3.5-9B\footnote{\url{https://huggingface.co/Qwen/Qwen3.5-9B}} and Qwen3.5-2B.
The first three run through their APIs at low reasoning effort.
Qwen3.5 has no effort levels, only thinking on or off; we serve the 9B and 2B models with vLLM and run them with thinking on, using the recommended sampling (temperature 1.0, top-p 0.95, top-k 20, presence penalty 1.5) and up to 16k output tokens, because at temperature 0 some calls repeated their reasoning until the output limit and ended without a tool call; we compare them with thinking off and greedy decoding in Section~\ref{sec:thinking}.

\paragraph{Metrics.}
The agent writes its answer as text in braces; Hits@1 lower-cases the answer and the gold names and removes spaces before matching.
We reproduce ToG's evaluation code with one change: an empty answer is wrong, whereas ToG's code counts it as a match.
An answer counts as correct if, after this normalization, it equals or contains a gold answer name, or is contained in one.
A brace list of several names thus matches as soon as it contains one gold name.
On SimpleQuestions, we also report Valid Hits@1 (Appendix~\ref{app:sq}).

\paragraph{ToG.}
We run ToG's released code and change only the endpoints, add bounded retries, and treat malformed SPARQL queries as empty results instead of stopping the run.
We keep its settings: beam width 3, depth 3, and temperature 0.4 for exploration and 0 for reasoning.
ToG makes 4 to 9 LLM calls per question.

\paragraph{SoG.}
SoG has published no code, so we reimplement it from its paper~\cite{sun2026sog}: one \tool{search} tool returns an entity's outgoing or incoming triples as a table, only the distinct relations when there are more than 50 rows and no relation filter is given, and at most 1{,}000 rows; the system prompt and tool description are copied verbatim, and the LLM keeps the whole search history in its prompt.
Where the paper leaves details open, we use five few-shot examples written by us in the paper's format from WebQSP and SimpleQuestions training questions; entity labels from English Freebase names, with ``-'' for unnamed nodes; rows sorted by relation and value before truncation; temperature 0; and a cap of 30 searches, after which the LLM is asked once to answer without the tool.
To stay within the query limits of our Virtuoso server, a filtered search runs one query per relation and looks up names in a separate query; this does not change the returned rows.

\paragraph{Protocol.}
Five CWQ questions of our agent ended with a provider error; they count as wrong.
Our agent abstains on 0.7\% (WebQSP) to 3.7\% (CWQ) of the questions.
A budget is exhausted on 0.9\% (WebQSP) to 6.5\% (CWQ) of the questions; the navigator then answers with the evidence it has.
At least one tool call fails on 7.9\% (SimpleQuestions) to 31.2\% (CWQ) of the questions, most often because the navigator cites a wrong reference, names a relation that was not offered, or names a field the rows do not have; each error costs one navigator call.

\section{SimpleQuestions Labels}
\label{app:sq}

\begin{table}[h]
  \centering
  \small
  \setlength{\tabcolsep}{4pt}
  \begin{tabular}{@{}lrcc@{}}
    \toprule
     & $n$ & Hits@1 & Valid \\
    \midrule
    \multicolumn{4}{@{}l}{\emph{Ours, by rows in the largest collection}} \\
    1--5 rows    & 710 & 90.0 & 90.3 \\
    6--100 rows  & 134 & 64.2 & 81.3 \\
    $>$100 rows  & 156 & 9.0  & 80.8 \\
    All          & 1{,}000 & 73.9 & 87.6 \\
    \midrule
    \multicolumn{4}{@{}l}{\emph{All systems, name-based Valid}} \\
    ToG (our run)     & 1{,}000 & 49.7 & 79.4 \\
    SoG (our reimpl.) & 1{,}000 & 73.9 & 92.5 \\
    Ours              & 1{,}000 & 73.9 & 92.0 \\
    \bottomrule
  \end{tabular}
  \caption{SimpleQuestions (\%). Hits@1 counts only the annotated object as correct; Valid counts any object of the gold subject and relation, e.g.\ any film directed by X. Top: the first returned entity must be valid. Bottom: the answer text must match the name of a valid object, which also applies to ToG and SoG, since they return only text.}
  \label{tab:sq}
\end{table}

Each SimpleQuestions question was written from one Freebase fact (subject, relation, object), and only that object is labelled as the answer~\cite{bordes2015simplequestions}.
Many questions, however, have several correct answers: for ``name a film directed by X'', every film directed by X is correct, but Hits@1 accepts only the one the annotator picked~\cite{petrochuk2018simplequestions}.
\emph{Valid Hits@1} instead counts an answer as correct if the first returned entity or value is an object of the gold subject and relation in our graph.
It checks only the first answer, so a list counts as correct if its first item is valid.
ToG and SoG return only text, so for them we use a name-based version: the answer counts as correct if it matches, with ToG's text match, the name of any object of the gold subject and relation.
This version is more lenient; we report it for our agent too, so that the three systems are compared under the same check.

We group the questions by the size of their largest collection, which for these one-relation questions is roughly the number of candidate answers (Table~\ref{tab:sq}, top).
With at most five candidates, the two metrics agree (90.0 vs.\ 90.3\%).
With more than 100 candidates, 80.8\% of the agent's answers are valid, but only 9.0\% are the annotated one.
On such questions, Hits@1 measures whether the agent guesses the annotator's choice among equally correct answers, not whether it answers the question.
Overall, 168 of the 261 questions that miss the annotated object (64\%) still have a valid answer.

The baselines are affected in the same way (Table~\ref{tab:sq}, bottom).
Under the name-based check, which we can apply to all three systems, our agent reaches 92.0\%, on par with SoG at 92.5\% ($p \geq 0.21$) and above ToG at 79.4\% (158 vs.\ 32 questions answered validly by only one of the two, $p < 0.001$).
The gap to SoG's published 84.8\% Hits@1 is therefore not specific to our agent; we attribute it to differences in the LLM and the graph, and possibly in how the answer is chosen among valid objects.

Valid Hits@1 does not check whether the question asks for one object or for all of them.
The agent returns a list on 37\% of the questions; 230 of its 739 Hits@1 hits are lists that contain the annotated name.

\section{Error Analysis}
\label{app:errors}

\begin{table}[t]
  \centering
  \small
  \setlength{\tabcolsep}{4pt}
  \begin{tabular}{@{}lr@{}}
    \toprule
    Cause & Questions \\
    \midrule
    Gold or question problem, or answer also correct & 26 \\
    Tool-call errors, loops or exhausted budget                 & 8 \\
    Wrong relation, or no path found                            & 6 \\
    Search stopped early or a condition not applied             & 6 \\
    Wrong \tool{filter} or \tool{select} decision               & 2 \\
    Decision-model request timed out                                 & 2 \\
    \bottomrule
  \end{tabular}
  \caption{Causes of 50 randomly sampled wrong answers on CWQ.}
  \label{tab:errors}
\end{table}

To see why answers are wrong, we inspected the traces of 50 randomly sampled wrong answers on CWQ (Table~\ref{tab:errors}).
The decision model's filtering and selection are rarely the cause: only 2 errors come from a wrong \tool{filter} or \tool{select} decision, and 2 from requests over collections of more than 1{,}000 rows that timed out; a wrong relation, which the decision model chooses unless the navigator names an alternative, accounts for at most 6 more.
About half are not errors of the agent: the gold answer does not answer the question as worded, the question is ambiguous, or the agent's answer is also correct, as for ``Which movie featuring Charlie Hunnam was released most recently?'', where the gold answer is Pacific Rim and the agent answered the later Crimson Peak.
This matches a recent audit, which found the gold answers of about half of the sampled WebQSP and CWQ questions to be incorrect~\cite{zhang2025kgqagen}.
The remaining errors come from the navigation: a wrong relation or a missed path, a search that stopped early or ignored a condition, and tool-call errors and loops, for instance when the navigator tries to intersect two collections, for which it has no tool.

\end{document}